\documentclass[12pt,article]{article}
\usepackage{times, amsmath, graphicx, setspace}
\title{Quantum Gravity, Yesterday and Today}
\author{Bryce DeWitt\footnote{The following text was found in Bryce's files without references and without an indication of its purpose} (1923-2004)\\ \vspace*{-.5cm}\small References and the name of a previously anonymous student have been supplied by\\ Cecile DeWitt and Brandon DiNunno}
\date{Dept. of Physics, University of Texas at Austin}
\begin{document}

\maketitle
 A paper that I published in 1967 \cite{1} began with these words: ``Almost as soon as quantum field theory was invented by Heisenberg, Pauli, Fock, Dirac, and Jordan, attempts were made to apply it to fields other than the electromagnetic field which had given it - and indeed quantum mechanics itself - birth. In 1930 Rosenfeld \cite{2} applied it to the gravitational field which, at the time, was still regarded as the \textit{other} great entity of Nature. Rosenfeld was the first to note some of the special technical difficulties involved in quantizing gravity and made some early attempts to develop general methods for handling them. As an application of his methods he computed the gravitational self energy of a photon in the lowest order of perturbation theory. He obtained a quadratically divergent result, confirming that the divergence malady of field theory, which had already been discovered in connection with the electron's electromagnetic self-energy, was widespread and deep seated. It is tempting, and perhaps no longer premature, to read into Rosenfeld's result a forecast that quantum gravidynamics was destined, from the very beginning, to be inextricably linked with the difficult issues lying at the theoretical foundations of particle physics.''

In 1967 such a forecast \textit{was} premature, and yet any thoughful person had to ask himself: What is the gravitational field doing there, in such splendid isolation? What if one simply dragged it forcibly into the mainstream of theoretical physics, and quantized it? In 1948 Julian Schwinger, my Ph.D. thesis advisor, gave me permission to reperform Rosenfeld's calculation, but in a manifestly gauge-invariant way, with the aim of showing that Rosenfeld's result implies merely a renormalization of charge rather than a nonvanishing photon mass. In fact, the one-loop result is nil, although to show this, using Schwinger's clumsy methods rather than Feynman's elegant diagrams, was not easy \cite{3}. 

Most of you can have no idea how hostile the physics community was, in those days, to persons who studied general relativity. It was worse than the hostility emanating from some quarters today toward the string-theory community. In the mid fifties Sam Goudsmidt, then Editor-in-Chief of the \textit{Physical Review}, let it be known that an editorial would soon appear saying that the \textit{Physical Review} and \textit{Physical Review Letters}\footnote[1]{Ed. S. Goudsmidt was also Managing Editor of \textit{Reviews of Modern Physics}} would no longer accept ``papers on gravitation or other fundamental theory.'' That this editorial did not appear was due to the behind-the-scenes efforts of John Wheeler. 

It was possible, in those days, to divide theoretical physicists into two camps, according to their view of the role of gravitation in physics, just as it is possible to divide mathematicians into two camps according to their answer to the question: Is mathematics \textit{there}, to be discovered, or is it a free invention of the human mind? With very few exceptions mathematicians of world class stature say that it is there. Those with limited horizons - the second raters - say that it is a free invention. Let me illustrate the situation in physics by the following anecdote:

In November 1949, at the Institute for Advanced Study (where I was already beginning to cast glances at Cecile), I met Pauli. I was hoping to spend some time as a postdoc at the ETH, so Pauli asked me what I was working on. I said I was trying to quantize the gravitational field. For many seconds he sat silent, alternately shaking and nodding his head (a nervous habit he had, affectionately known as die Paulibewegung). He finally said ``That is a very important problem - but it will take someone really smart!''

Neither Pauli nor Schwinger had limited horizons, nor had Feynman, who began to think about quantum gravity after attending a conference on the role of gravitation in physics organized by Cecile in January 1957 \cite{4}\cite{5}. Feynman's thinking culminated in his discovery of ghosts, which he announced at a gravity conference in Warsaw in 1961 \cite{6}. At that time he knew how to incorporate ghosts only into one-loop diagrams. The two-loop case, with its overlapping subgraphs, was unravelled by me in 1964 \cite{7}, by methods that would clearly give me the answer in any order. By the end of 1965 I was able to express the rules for ghosts to all orders in terms of a functional integral that could easily be shown to be invariant under deformations in gauge-breaking terms. These rules were written up and submitted to the Physical Review in 1966 \cite{8}.

Here I have to backtrack and describe the situation outside the physics community. In 1955 I received a letter from the Glenn L. Martin Aircraft Company which began with the words ``It occured to us a number of years ago that our company was vitally interested in gravity ....'' They were looking for physicists who could build an antigravity device, and turned to me because I had won first prize in a Gravity Research Foundation essay contest. In those days, only a decade from Hiroshima and Nagasaki and only two and a half years from the hydrogen bomb, physicists were viewed as gods who could do anything. Athough I did not accept the Glenn L. Martin offer I did profit from the research-grant environment. It was the Air Force who supported my research on quantum gravity as well as postdocs such sa Peter Higgs, Heinz Pagels and Ryoyu Utiyama whom I had invited.

But by 1966 the military realized that they weren't going to get magical results from gravity research, and my Air Force grant was terminated. This meant that I was unable to pay the page charges that the \textit{Physical Review} was levying in those days, with the result that the paper I submitted in 1966 was not processed and published until over a year later. Nowadays, of course, everyone goes on the web and receives instantaneous exposure.

The story of ghosts is not the only feature of early quantum gravity research. In 1949 Peter Bergmann \cite{9} began to look for a quantum analog of the 1938 work of Einstein, Infeld and Hoffmann \cite{10} on the motions of singularities in the gravitational field. Since these motions follow from the gravitational field equations alone, without divergent quantities or such concepts as self-mass appearing at any time, Bergmann reasoned that this might be a way to avoid field-theory divergences altogether. Nowadays we regard Bergmann's vision as rather archaic. But it was a vision that was shared by Dirac, who always viewed particles (e.g. electrons) as somewhat different from fields, and even by Feynman, who obtained his famous propagator by carrying out a complex Laplace transform on a heat kernel obtained from a path integral over explicit particle trajectories, both forwards and backwards in time \cite{11}. 

Feynman's conception at least had the merit of being manifestly covariant. But Bergmann had to approach the Einstein-Infeld-Hoffmann picture from the canonical side, singling out the time for special treatment. Although Bergmann's vision never really got off the ground, intensive work was carried out in those years on canonical quantum gravity, culminating in an equation that bears my name along with that of John Wheeler \cite{12} who was the real driving force. Research on the consequences of this equation continues to this day, stimulated by work of Abhay Ashtekar \cite{13}, and some of it is quite elegant. But apart from some apparently important results on so-called ``spin foams'' \cite{RGB} I tend to regard the work as misplaced. Although WKB approximations to solutions of the equation may legitimately be used for such purposes as calculating quantum fluctuations in the early universe \cite{14}, and although the equation forces physicists to think about a wave function for the whole universe and to confront Everett's many-world view of quantum mechanics \cite{15}, the equation, at least in its original form \cite{12}, cannot serve as the \textit{definition} of quantum gravity. Aside from the fact that it violates the very spirit of general relativity by singling out spacelike hypersurfaces for special treatment, it can be shown not to be derivable, except approximately, from a functional integral \cite{12}. For me the functional integral must be the starting point. 

I cannot leave the Einstein-Infeld-Hoffmann-Bergmann-Dirac-Feynman story without mentioning one difference between particles (specifically fermions) and fields (specifically bosonic fields) that raises an issue of terminology. Despite the fact that mathematicians have found the Dirac operator to be more fundamental than the Laplacian \cite{RB} (at least in index theory) fermions, unlike bosons, cannot even be introduced into spacetime unless spacetime satisfies certain Strefel-Whitney conditions allowing the introduction of a spin or pin bundle \cite{RC}. By the two-to-one homomorphism from the spin (or pin) group to the Lorentz group, such a bundle defines a Lorentz-frame bundle. A local trivialization of the Lorentz-frame bundles defines a field of local Lorentz frames. These were invented by Elie Cartan who called them \textit{reperes mobiles}. There is a well documented exchange of letters \cite{EINSTEIN} in which Cartan tried in vain to get Einstein to understand the value of using \textit{reperes mobiles} \cite{RD}. When Pauli and others finally got the point the jargon of a German speaking in-group took over, resulting in such barbarisms as \textit{vielbeine} or \textit{bein} rotations. In my view it is inapporpriate to use German jargon for something invented by a Frenchman. As for English jargon, ``frame'' is at least the correct translation of ``repere''

After the rules for ghosts to all orders were settled, progress was rapid. `t Hooft and Veltmann invented dimensional regularization \cite{b}, which kept gauge invariance intact and allowed Zinn-Justin and others to show in detail the renormalizability of the Yang-Mills field minimally coupled to other renormalizable fields \cite{RE}. Zinn-Justin's proof of the gauge invariance of counter terms is applicable to the gravitational field, but there the best one can hope for is a low energy effective theory, obtained by minimal subtraction, order by order. Although efforts \cite{c} to make the gravitational field serve as its own cut-off, in some nonpertubative way, had been undertaken several times in earlier years, none of these efforts panned out. 

The modern history of quantum gravity begins with Stephen Hawking and his discovery of the thermal radiation emitted from black holes formed by gravitational collapse \cite{RF}. The thermality of the radiation allows one to assign a temperature and an entropy to a black hole \cite{Bek}. The entropy of a solar-mass black hole turns out to be fantastically large, twenty orders of magnitude larger than the entropy of the sun itself. And the disparity is even greater for larger masses. This suggests that the entropy of a black hole is the maximum entropy that any object with the same size and mass can have, an idea that has spurred many attempts to compute the entropy from first principles by summing over putative internal states of the black hole. Perhaps the most successful of these efforts have been string theory computations for certain extremal black holes \cite{RGA} and the spin-foam computations for Schwarzschild black holes \cite{RGB}.

In viewing string theory one is struck by how completely the tables have been turned in fifty years. Gravity was once viewed as a kind of innocuous background, certainly irrelevant to quantum field theory. Today gravity plays a central role. Its existence \textit{justifies} string theory! There is a saying in English: ``You can't make a silk purse out of a sow's ear.'' In the early seventies string theory was a sow's ear. Nobody took it seriously as a fundamental theory. Then it was discovered that strings carry massless spin-two modes \cite{RH}. So, in the early eighties, the picture was turned upside down. String theory suddenly \textit{needed} gravity, as well as a host of other things that may or may not be there. Seen from this point of view string theory is a silk purse. I shall end my talk by mentioning just two things that, from a nonspecialist's point of view, make it look rather pretty. 

In 1963 I gave [Walter G. Wesley] a student of mine the problem of computing the cross section for a graviton-graviton scattering in tree approximation, for his Ph.D. thesis \cite{blank}. The relevant diagrams are these:
%%%%%%%%%%%%%%%%%%%%%%%%%%%%%%%%%%
\begin{center}
\includegraphics[scale=.7]{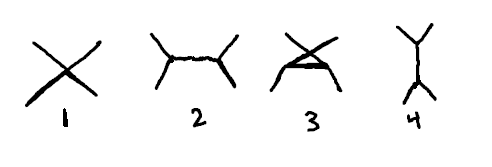}
\end{center}
%%%%%%%%%%%%%%%%%%%%%%%%%%%%%%%%%%
Given the fact that the vertex function in diagram 1 contains over 175 terms and that the vertex functions in the remaining diagrams each contain 11 terms, leading to over 500 terms in all, you can see that this was not a trivial calculation, in the days before computers with algebraic manipulation capacities were available. And yet the final results were ridiculously simple. The cross section for scattering in the center-of-mass frame, of gravitons having opposite helicities, is
\begin{center}
$d\sigma/d\Omega = 4G^2E^2\cos^{12}\frac{1}{2}\theta/\sin^4\frac{1}{2}\theta$
\end{center}
where $G$ is the gravity constant and $E$ is the energy \cite{blank}.

In string theory there is only one diagram, namely
%%%%%%%%%%%%%%%%%%%%%%%%%%%%%%%%%
\begin{center}
\includegraphics[scale = .7]{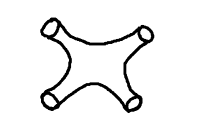}
\end{center}
%%%%%%%%%%%%%%%%%%%%%%%%%%%%%%%%%
and its contribution to the graviton-graviton amplitude is relatively easy to compute, giving the same result as that obtained by my student.

The other ``pretty'' feature of string theory concerns the topological transitions. In conventional quantum gravity topological transitions are impossible. I say this despite occasional efforts that have been made in the past to sum ``amplitudes'' for different spacetime topologies in ``Euclidean quantum gravity,'' ``Euclidean'' being chosen to avoid the singularities necessarily accompanying changes of spatial topology in Lorentzian manifolds. In the first place, Euclidean quantum gravity simply does not exist, because the Euclidean action is not bounded from below. Moreover, there is no classification of topological transitions analogous to the homotopy classification of paths discovered by Cecile and her student Laidlaw \cite{A}, which enables one to assign phases to the contributions to path integrals from different homotopy classes, based on the one-dimensional representation of the fundamental group. Cecile's methods are directly applicable to the Yang-Mills field, for which a precise homotopy classification exists. But no group analogous to $\pi_1$ exists for the topological analysis of Lorentzian quantum gravity.

In string theory, on the other hand, one finds that strings can live perfectly well on \textit{orbifolds}, which constitute a certain generalization of manifolds. With orbifolds, even Lorentzian orbifolds \cite{B}, topological transitions become possible. Therefore John Wheeler's forty-year-old vision of spacetime foam may be a reality \cite{C}.
\singlespace

\end{document}